\documentclass[a4paper,11pt]{article}
\pdfoutput=1 % if your are submitting a pdflatex (i.e. if you have
\usepackage{jinstpub} % for details on the use of the package, please
\title{\boldmath A fully-active fine-grained detector with three readout views}

\author[a]{A.~Blondel,}
\author[a]{F.~Cadoux,}
\author[b]{S.~Fedotov,}
\author[b]{M.~Khabibullin,}
\author[b]{A.~Khotjantsev,}
\author[a]{A.~Korzenev,}
\author[b]{A.~Kostin,}
\author[b,d,e]{Y.~Kudenko,}
\author[c]{A.~Longhin,}
\author[b]{A.~Mefodiev,}
\author[a]{P.~Mermod,}
\author[b]{O.~Mineev,}
\author[a]{E.~Noah,}
\author[a,*,1]{D.~Sgalaberna,\note{Corresponding author.}}
\author[b]{A.~Smirnov,}
\author[b]{N.~Yershov}

\affiliation[a]{University of Geneva, Geneva, Switzerland} % (now at CERN)}
\affiliation[b]{Institute for Nuclear Research  of the Russian Academy of Sciences, Moscow, Russia}
\affiliation[c]{INFN Padova, Italy}
\affiliation[d]{Moscow Institute of Physics and Technology,  Moscow Region, Russia}
\affiliation[e]{Moscow Engineering Physics Institute (MEPhI), Moscow, Russia}
\affiliation[*]{Now at CERN}

\emailAdd{davide.sgalaberna@cern.ch}

\abstract{
This paper describes a novel idea of a fine-grained fully-active plastic scintillator detector 
made of many optically independent $1\times1\times1~\text{cm}^3$ cubes 
with readout on three orthogonal projections by wavelength shifting fibers. 
The original purpose of this detector is to serve as an active neutrino target for the detection,
 measurement and identification of the final state particles 
 down to 
 a few tenths MeV kinetic energies.
The three readout views 
as well as the fine granularity
ensure powerful localization and measurement of the deposited energy combined with good timing properties and isotropic acceptance. The possible application as a new active target for the T2K near detector, initial simulation studies and R\&D test results are reported.
}


\arxivnumber{1707.01785} % only if you have one

\begin{document}
\maketitle 
\flushbottom

\section{Introduction}
\label{sec:intro}

Plastic scintillator material is very commonly used in high-energy 
physics and astroparticle physics, providing detectors with good timing 
properties (often below 1~ns resolution) 
and measurement of the deposited 
energy. The advent of scintillating fibers and/or wavelength shifting 
fiber readout allows for very flexible geometrical and tracking properties. 
Neutrino  experiments have used scintillators quite systematically, recent 
examples being the MINOS~\cite{MINOS}, Miner$\nu$a~\cite{MINERvA} 
and the near detectors suite (ND280) of T2K~\cite{Abe:2011ks,Abe:2016tii}. In the case of ND280, the 
readout with silicon photo-multipliers was applied systematically to 60'000 channels,
allowing a more compact geometry to be achieved.

In the above examples, narrow plastic scintillator bars are disposed
perpendicularly to the neutrino beam direction, a geometry that is 
suitable for neutrino interactions in beams of energy above typically a few GeV, for which  the 
leading final state particles are predominantly emitted in the forward 
direction. For lower neutrino energies, the final state lepton of the 
neutrino charged current interactions is emitted more isotropically. In 
that case it is interesting to aim at a more isotropic geometry. At all energies, nevertheless,  
nuclear effects commensurate to the binding energy of the 
Carbon nucleus or its Fermi Momentum, occurring either in the initial
state or in the final state of the neutrino interaction, affect the 
energy balance and the energy reconstruction \cite{t2k_longpaper_nuecc1pi}. 
It is thus important 
to be able to study the effect of nuclear activity by locating the energy 
deposited by additional nucleons originating from the interaction or 
resulting from nuclear breakup. In this context the natural granularity scale 
is around 1~cm, corresponding to  the range in plastic of protons 
with momentum commensurate  with the Fermi motion 
%of 222 MeV/c. 
of about 220 MeV/c. Another example of application of a more isotropic geometry could be the 
astroparticle physics experiments where the detector
orbits around the Earth and can detect particles produced by several
different sources and coming from any direction.

In case of a detector with scintillator bars disposed perpendicularly 
to the beam axis (hereinafter this axis referred to as $Z$), acceptance and resolution are 
highly direction-dependent: a particle traveling along a single scintillator bar
%for a large fraction of its path 
cannot be 
%well 
tracked 
and the momentum cannot be defined. Furthermore, in a realistic 
situation several tracks can be produced and it often happens that 
the energy deposited cannot be uniquely identified 
and assigned to a particular spatial direction. 
In this case a three-dimensional readout of the signals will 
ensure a more isotropic acceptance and reconstruction. 

In this paper, concentrating on the case of the project of the ND280 detector upgrade, 
we present an attempt at such a 3-D design, keeping in mind the need to have the number of channels 
to a reasonable value
and potential applications to other fields in physics.
%There are probably many other applications in particle physics or elsewhere. 

This article is organized as follows: 
the detector concept and design is described in Sec.~\ref{sec:design};
in Sec.~\ref{sec:test} the measurements performed on a small prototype with cosmic particles is shown;
finally in Sec.~\ref{sec:simulation} the simulation results of the proposed detector are described.

\section{The design of the detector}
\label{sec:design}

%All the neutrino oscillation parameters \cite{{Maki:1962mu}} have been measured 
%except the CP violating phase that, if equal to 0 or $\pi$, introduce
%CP violation in the leptonic sector.
The goal of the currently  running long baseline experiments, T2K and NO$\nu$A \cite{NOvA},
is to measure the CP violating phase in the neutrino sector, by measuring neutrino appearance phenomena, such as the 
$\nu_{\mu} \rightarrow  \nu_{e}$ and $\bar{\nu}_{\mu} \rightarrow  \bar{\nu}_{e}$ transitions. 
To this effect, one compares the neutrino event rate at a near detector, 
before oscillations occur,
with the neutrino event rate at the far detector,
whose position, in the case of T2K and NO$\nu$A, is located near the oscillation maximum.
On a longer time scale,  new experiments, Hyper-K \cite{t2hk} and DUNE \cite{dune},
will start searching for CP violation with much larger data sample.
For this reason it is timely to develop  near detector designs in which 
%all necessary
as much as possible 
information is acquired, both by establishing  the rate and flavour of neutrino interaction events and by understanding the measurement of neutrino energy -- since neutrino energy is the quantity which governs the neutrino oscillations. This latter point requires a detailed knowledge of neutrino interactions. Furthermore, the possible differences between electron- vs muon- (anti)neutrino cross sections are essential for the precise measurement of the appearance oscillation phenomenon.   
Several experiments 
\cite{Abe:2011ks,MINERvA,WAGASCI,NOvA}
are currently measuring the neutrino interaction cross sections with scintillator detectors. 
%with increasing  precision, 
However a dedicated effort is required to address the specific needs of the neutrino oscillation program.  
In the case of T2K, these have been spelled out in the T2K ND280 upgrade program: 
\begin{itemize}
\item 
the near detector measurements must cover the full polar angle range for the final state lepton with a  well understood acceptance; 
\item 
the near detector must be capable of measuring (at least the ratio of) electron and muon neutrino cross sections;
\item
the near detector should be able to address the issue of nuclear effects and their impact on energy reconstruction.
\end{itemize}
 
The detector must also be fully active and the amount of dead material
must be minimized, in order to detect all the energy released by the produced particles
and reconstruct with precision the energy of the interacting neutrino.
Furthermore, 
in an experiment like T2K, Hyper-K and NO$\nu$A it becomes
very important 
%to measure the neutrino interactions in nuclei very similar to the 
%target ones at the far detector, 
to have very similar nuclear targets at the near and far detectors, 
for instance $\text{H}_2\text{O}$ at T2K (Hyper-K) and
liquid scintillator at NO$\nu$A.
Organic scintillators, $^{12}{\text C}$-based, fulfill as much as possible this requirement. 
A good compromise would be given by a fine granularity detector, $\sim1~\text{cm}$,
the range of a proton with the Fermi momentum,
with a good acceptance 
over the full solid angle.

An interesting solution is a full 3D detector \cite{calocube}, 
%where each hit of the detector can be uniquely assigned to a track, 
that would solve the tracking ambiguity issues.
However given the combination of large mass and very fine granularity, 
a 
%extremely large
prohibitively large
number of readout channels
($O(1\text{M})$, since it scales with the detector volume) 
would be required if one would read out individually cm-size cubes, leading to high costs and a large amount of dead material.
We propose here 
an alternative solution 
%that could achieve the requirements
%described above but 
with 
more acceptable
costs and 
less
dead material inside the neutrino target.

\begin{figure}
\begin{center}
\includegraphics[height=5cm]{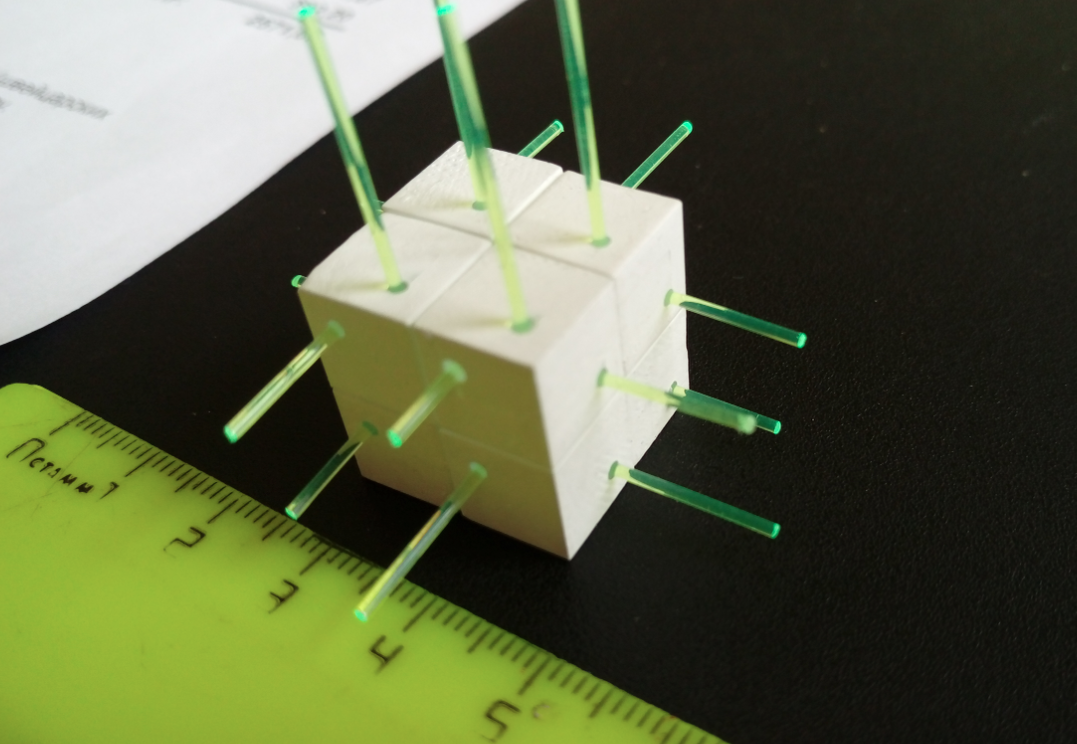}\\[6pt]
\includegraphics[height=5cm]{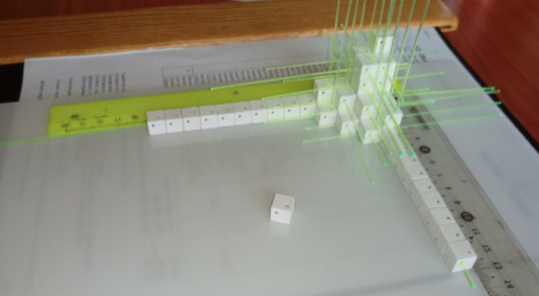}\\[6pt]
\includegraphics[height=5cm]{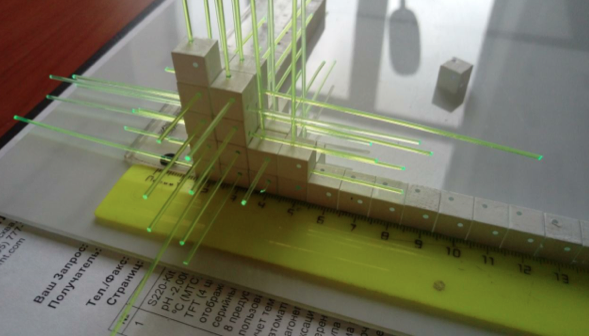}\\[6pt]
\caption{\label{prototype}The picture of a small prototype is shown.
Several cubes of extruded plastic scintillator with three WLS fibers 
inserted in the three holes are assembled.
The size of each cube is $1\times1\times1~\text{cm}^3$.}
\end{center}
\end{figure}

The proposed detector consists of many cubes of extruded scintillator, each one covered by a reflector and read out along three orthogonal directions by wavelength shifting fibers. 
The chosen scintillator is a composition of  a polystyrene doped with 1.5\% of paraterphenyl (PTP) and 0.01\% of POPOP. 
The cubes produced by Uniplast, a company in Vladimir, Russia,  
are covered by a $\sim 50~\mu \text{m}$ thick 
%chemical reflector. 
diffusing layer.
%\textcolor{red}{
%COMMENT 2:
%Is the micropore deposit over the scintillator cubes (maybe more adequately described as a diffuser) more or less efficient in allowing light to reach the fibers than an Aluminum reflecting coating? Do you have any data on this matter? Totally reflective coating would eliminate the crosstalk (see the next point).
%}
%\textcolor{blue}{
%NEW TEXT:
Depending on the physics case, different technologies can be used for the reflector
to constrain the scintillation light inside the cube.
In order to provide good light yield and uniformity in the neutrino target nuclei,
the reflector is obtained by etching the scintillator surface in a chemical  agent that results in the formation of a white micropore deposit, acting as a diffuser, over a polystyrene surface~\cite{Kudenko:2001qj}.
%}
Each cube has three orthogonal cylindrical holes of 1.5~mm diameter drilled  along X, Y and Z axes. 
Three 1.0~mm  wavelength shifting (WLS) fibers, multi-clad Kuraray Y11, are inserted through the holes. The ideal size of each cube is $1\times1\times1~\text{cm}^3$, providing the required  fine granularity.  
%\textcolor{red}{
%COMMENT 5:
%Coordinate system: I suggest you introduce your coordinate system at page 3, where you give the size of your proposed detector, specifying also the direction of the neutrino beam in this system. Then the coordinates you mention in the second paragraph of Section 4 would be clearer.
%}
%\textcolor{blue}{
%NEW TEXT:
The axes $X$, $Y$ and $Z$ define respectively the width, the height and the length of the detector
and the neutrino beam direction is supposed to be centered along the $Z$ axis.
%}
We consider a detector of the size of $1.8\times0.6\times2.0~\text{m}^3$,
i.e. about 2 tons,
that would correspond to approximately 59k readout channels. 
Should this number prove to be too high, the size of the cubes could be increased for instance up to 
$2\times2\times2~\text{cm}^3$, corresponding to only 15k readout channels. 
For a given detector size, 
the number of cubes is inversely proportional to the third power of the cube size, 
while the number of channels scales as the inverse of the square. %of the cube size. 
A picture of a small prototype is shown in Figure~\ref{prototype}.
The parameters of the detector and scintillator cubes are shown
in Table~\ref{Table:parameters_detector} and \ref{Table:parameters_cube}.

The power of such a detector is given by 
the three views that would provide a $4\pi$ acceptance
and solve the tracking ambiguity efficiently.
%
%\textcolor{red}{
%COMMENT 1:
%The most important item of information that I would like you to add is the following:
%Reading out individual hits with the orthogonal 3D fiber system you describe gives three sets of N coordinates in which the association of each x with an other y and z is not obvious. For example, see the 2D case of two hits with coordinates x1,y1 and x2,y2: in lack of ad- ditional information, one cannot exclude that the hits were at x1,y2 and x2,y1. The number of possible combinations rapidly grows with N. Fortunately, real tracks provide additional information. Linking to- gether hits to form tracks should resolve most ambiguities.
%You do not address this issue. You should address it in Section 4, on simulations, or before. \\
%}
%\textcolor{blue}{
%NEW TEXT:
Thanks to this novel configuration we expect it will be possible to associate the particle hits to the right track
for most of the neutrino events.
The third view will help solving the ambiguity providing a great improvement compared to the two-views detectors,
where different tracks can share the same hits in the 2D projection.
This will be possible also thanks to the fact that O(1~GeV) neutrino interaction events 
 are usually low multiplicity. 
Furthermore the track reconstruction itself, where hits are linked together, will help.
Some ambiguities are still expected for hits close to the neutrino interaction vertex, 
within a few cms, where some low-momentum nucleons could be ejected by the nucleus
within a very close range.  
In this case, thanks to the fine granularity,
this detector can provide a precise calorimetry of the energy released,
%helping to infer the number of the released nucleons
improving the final reconstruction of the neutrino energy.
%}

%
%Neutrino experiments at the T2K beam energy (typically 650 MeV) have a relatively low multiplicity
%and a full 3D detector is not required for a good tracking.  
At the T2K most probable neutrino energy, typically 600~MeV, the event 
multiplicity is low and a full cell-by-cell readout, which would be 
prohibitive both in terms of number of readout channels and added 
passive material, 
is 
%probably 
not necessary. 
Our first investigations 
show that a system with a three-view readout 
looks appropriate.
%is already excellent.
%
The fine granularity would allow to measure protons with 
momenta 
%close 
down
to 300~MeV/c.
None of the above properties could be achieved with a scintillator bar detector. 
In addition the light is enclosed within the cube and the light yield is expected to be higher than in standard scintillator bars.
In this configuration, the energy deposited by a given ionizing 
particle is simultaneously collected by three WLS fibers, instead of only one,
improving pattern recognition and the light yield.
%timing and the particle identification (PID) by dE/dx.
Furthermore, in order to obtain three views,
the energy deposited by a particle in a single cube is enough.
This is a tremendous
advantage compared to scintillator bar detectors
where two views are provided by a particle depositing energy in at least
two different bars.

This detector is quite useful to distinguish pions and muons
%, on the other hand, 
from protons and electrons. 
In addition it is essential for separating the electrons, 
which are a manifestation of 
the electron neutrino interactions,  from the $\gamma \rightarrow e^+ e^-$ background. 
%
%Since neutrino interactions are usually low multiplicity, in particular at the energies below 1 GeV,
%this detector would provide performances similar to a 3D detector but, at the same time, 
%drastically reduce the number of channels, that would scale with $3 \times \text{N}^2$ instead of $\text{N}^3$.

This detector could be also used to detect neutrons produced in the neutrino interaction.
Indeed an additional coating of either Gadolinium or Lithium could be applied on 
the surface of each cube, similarly to what has been done for the SOLID experiment \cite{solid}:
the neutrons are captured after being thermalized and photons, 
delayed with respect to the lepton produced by the interaction, are released. 
As already mentioned above, the conceived geometry could be useful also 
for astroparticle physics experiments, thanks to the full solid angle acceptance.

% NEW SENTENCE: DO WE LEAVE IT???
Different configurations of the detector could be obtained in order to fulfill different requirements, 
such as an improved angular resolution. 
In this case half of the plastic cubes could be replaced with a very low density material,
e.g. AIREX 
(with a density of $0.06 \text{ g/cm}^3$, about 
6\% density of that of plastic scintillator) 
allowing tracks to travel for a longer
distance and improve their separation.
These aspects are not addressed here.

\begin{table}[!tbp]
  \centering
  \caption{
  Main parameters of the proposed detector of the size of $1.8 \times 0.6 \times 2.0~\text{m}^3$.
  }\label{Table:parameters_detector}
  \begin{tabular}{c|c|c}
   \hline
   \hline
   Parameter      				& Cube edge: 1~cm		& Cube edge: 2~cm		 \\
    \hline
    \hline
    \# of cubes					& 2.16M				& 270k	\\
    \hline
    \# of channels				& 58.8k				& 14.7k			\\
    \hline
    \hline
  \end{tabular}
 \end{table}

\begin{table}[!tbp]
  \centering
  \caption{
  Main parameters of each scintillator cube with 1~cm edge.
  }\label{Table:parameters_cube}
  \begin{tabular}{c|c}
   \hline
   \hline
   Parameter      				& Value		 \\
    \hline
    \hline
    Coating thickness				& $50~\mu m$			\\
    \hline
    Hole diameter				& 1.5~mm			\\
    \hline	
    WLS fiber diameter			& 1.0~mm 			\\
    \hline
    \hline
  \end{tabular}
 \end{table}

\section{Measurements}
\label{sec:test}
\begin{figure}[ht]
\centering % \begin{center}/\end{center} takes some additional vertical space
\includegraphics[width=12cm]{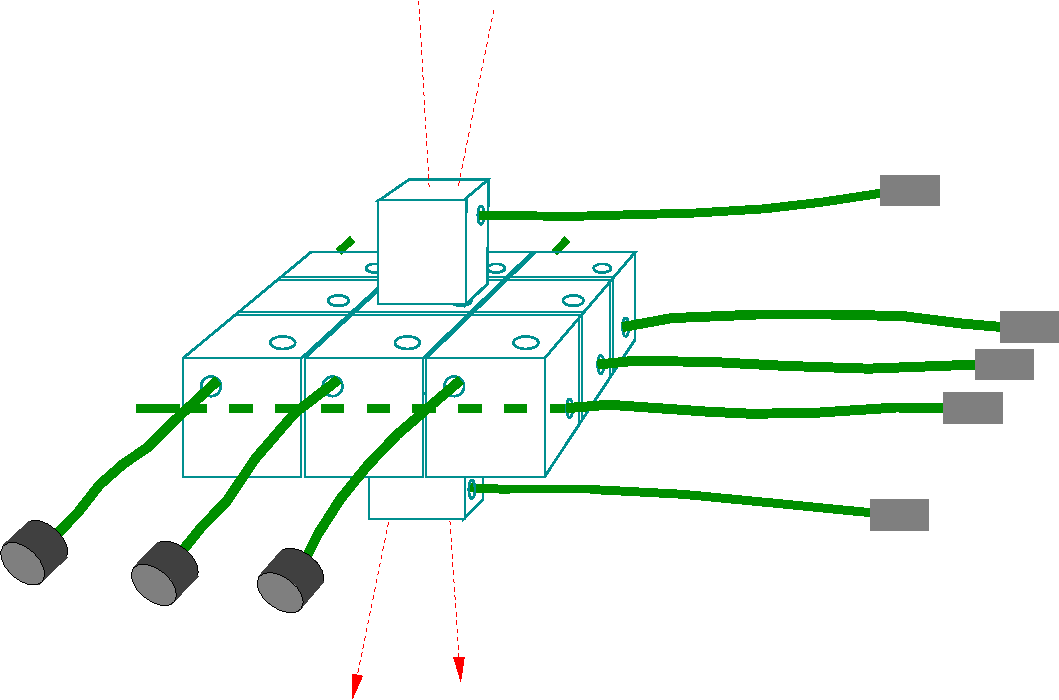} 
\caption{\label{fig:setup} Test bench for study of the parameters of scintillator cubes  using cosmic muons. The tested array comprised nine $1\times1\times1$ cm$^3$ cubes. Two trigger counters, each of a $8\times 8$ mm$^2$ cross section, 
were located above and below the tested cubes. }
\end{figure}

In order to study the performance of the scintillator cubes, we carried out measurements of the light yield produced from cosmic ray muons using a small plastic counter. The test bench for detector measurements is shown in Figure~\ref{fig:setup}.
The ionization area within the tested cube was localized to a $8\times 8$ mm$^2$ spot defined by the trigger counter size.
The readout of the scintillation light from each cube was provided by a 1.3 m long double-clad Kuraray Y11 WLS fiber  
coupled at one end to a photosensor,  
a Hamamatsu multi-pixel photodiode (MPPC).  
The sensor S12571-025C~\cite{hamamatsu}  consists of an array of 1600 independent $25\times
25$ $\mu$m avalanche photodiodes (pixels) operating in Geiger mode. The MPPC sensitive area is   $1\times 1$ mm$^2$. The photo detection efficiency of this MPPC  is about 35\% (3.5~V overvoltage) for green light of 520 nm as emitted by  a Y11 fiber.      
% Description of the electronics
Signals from MPPCs were amplified by a custom-made
preamplifier with a gain of 20, then sent to the 5 GHz sampling digitizer CAEN DT5742 with 12-bit resolution. 
The signal charge was calculated as an area of signal waveform normalized to number of photoelectrons. 
The signal timing was obtained at the 10\% fraction of the signal amplitude.
The position of all cubes along the WLS fibers inserted in the holes was fixed at the distance of 1 m from the photosensor. 
In order to increase the light yield, the far end of the fiber was covered by a teflon tape.

\begin{figure}[ht]
\centering % \begin{center}/\end{center} takes some additional vertical space
\includegraphics[width=9cm]{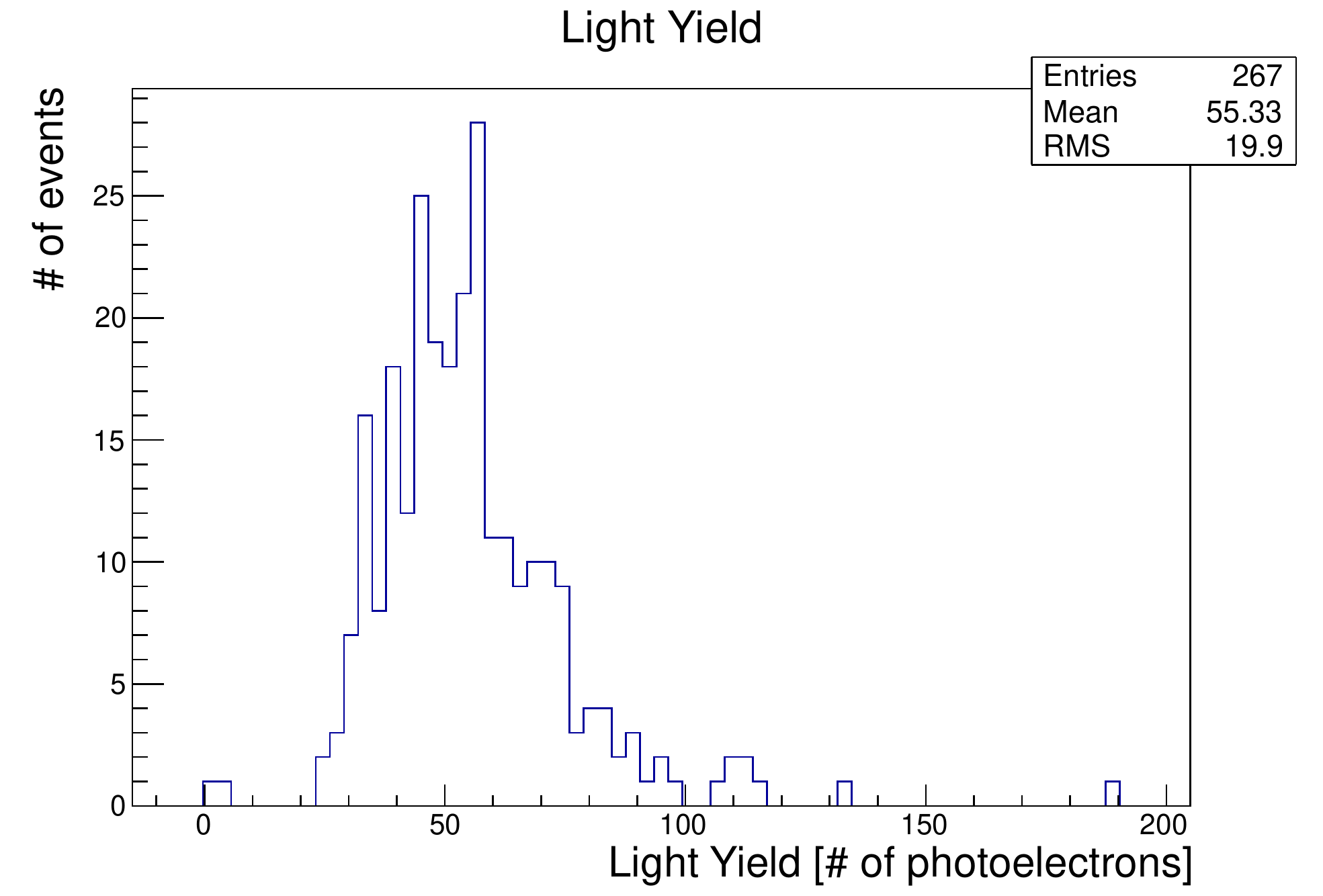} 
\caption{\label{fig:ly} The light yield of a scintillator cube per a minimum ionizing particle with one WLS fiber measured at a distance of 1 m from the MPPC. }
\end{figure}

The light yield of one scintillator cube  in photoelectrons (p.e.) per minimum ionizing particle (MIP) obtained with one fiber is plotted in Figure~\ref{fig:ly}. 
%The light yield of about 
%55 p.e./MIP % no gaussian fit  
%At a distance of 1 m from the MPPC about 55 p.e./MIP were measured for one fiber.
About 55 p.e./MIP were measured with a 1~m long WLS fiber.
%Following this result, one can expect the light yield  of 
%more than
%150 p.e./MIP
%in one cube  for the sum of three fibers. 
For a 
%maximum 
fiber length of 2 m the estimated light yield is expected to be about 35 p.e. according to the attenuation length of Y11 for green light~\cite{Mineev:2011xp}.

\begin{figure}[ht]
\centering % \begin{center}/\end{center} takes some additional vertical space
\includegraphics[width=9cm]{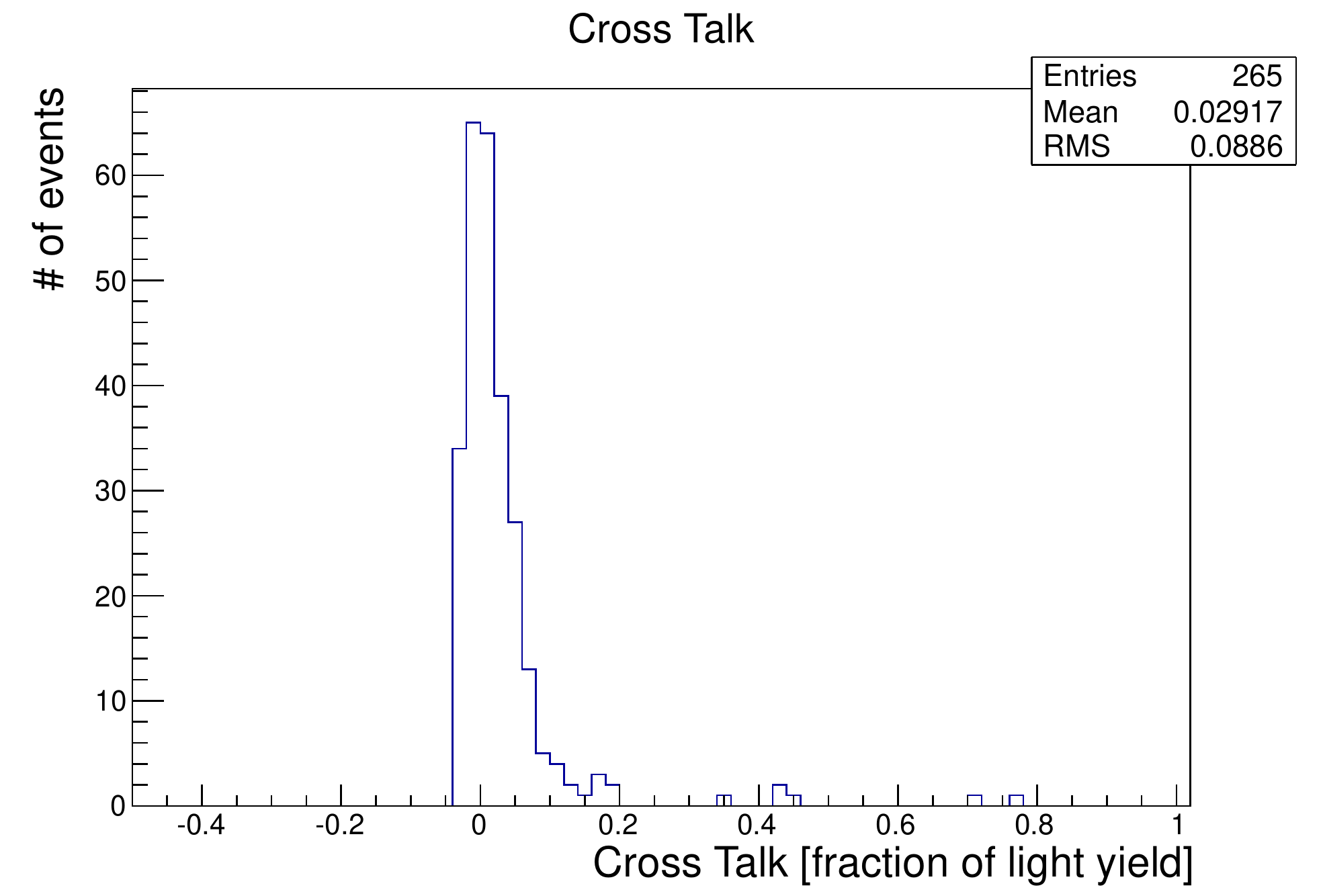} 
\caption{\label{fig:crosstalk}The ratio of the signal in the adjacent cube  to the  signal in the fired central cube. Negative values  are caused by subtracting the average dark noise contribution
on the  event-by-event basis.
%\textcolor{red}{
%COMMENT 4:
%On the caption of Fig. 4: delta-rays seem to me a more likely cause for the larger energy depositions (more than 0.2). If you agree, please mention this cause.
%}
%\textcolor{blue}{
%NEW TEXT:
Values higher than 0.2 are likely due not to light cross talk but,
in large part, to delta-rays
or multiple scattering of the cosmic ray 
in the scintillator cube that produce signal in the adjacent cubes. 
These entries are taken into account in the final cross talk calculation.
%}
%\textcolor{red}{
%COMMENT 3:
%A clarification on the measurement of crosstalk in Fig. 4: please explain how you obtain the 0.029 crosstalk fraction from the distribution shown in the figure. There is more than one way to account for the negative- amplitude noise events.
%}
%\textcolor{blue}{
%NEW TEXT:
The cross-talk is estimated to be 2.9\%, that corresponds to the mean of the shown distribution.
%}
}
\end{figure}

Since the white chemical reflector 
%does not provide perfect light reflection, the
does not fully contain the scintillation light, its
leak of the scintillating light from one cube to neighboring ones was investigated. The cross-talk was measured between the fired  central cube and adjacent cubes which surround the central one as shown in Figure~\ref{fig:setup}. The cross-talk is defined as the ratio of the light yield in  an adjacent cube to the  signal in the central cube. 
%The signal in the adjacent cube was corrected to the MPPC dark noise which was measured simultaneously during this test. 
The MPPC dark noise was measured simultaneously during the test and subtracted from the signal in the adjacent cube. 
Figure~\ref{fig:crosstalk}
shows that 
%about 2\% 
on average less than 3\%
of scintillating light penetrates from one cube to another.

\begin{figure}[ht]
\centering 
\includegraphics[width=7.5cm]{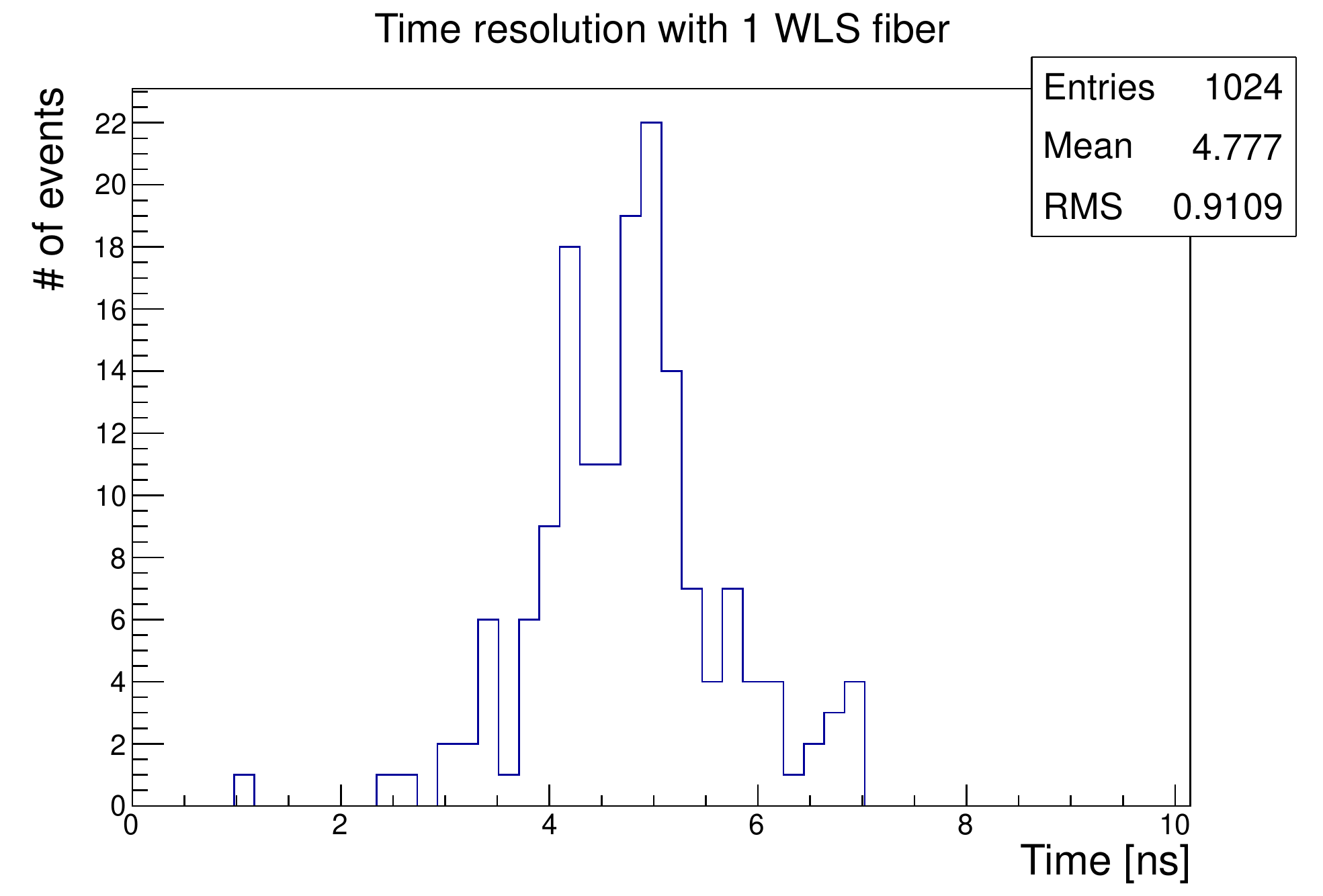} 
\includegraphics[width=7.5cm]{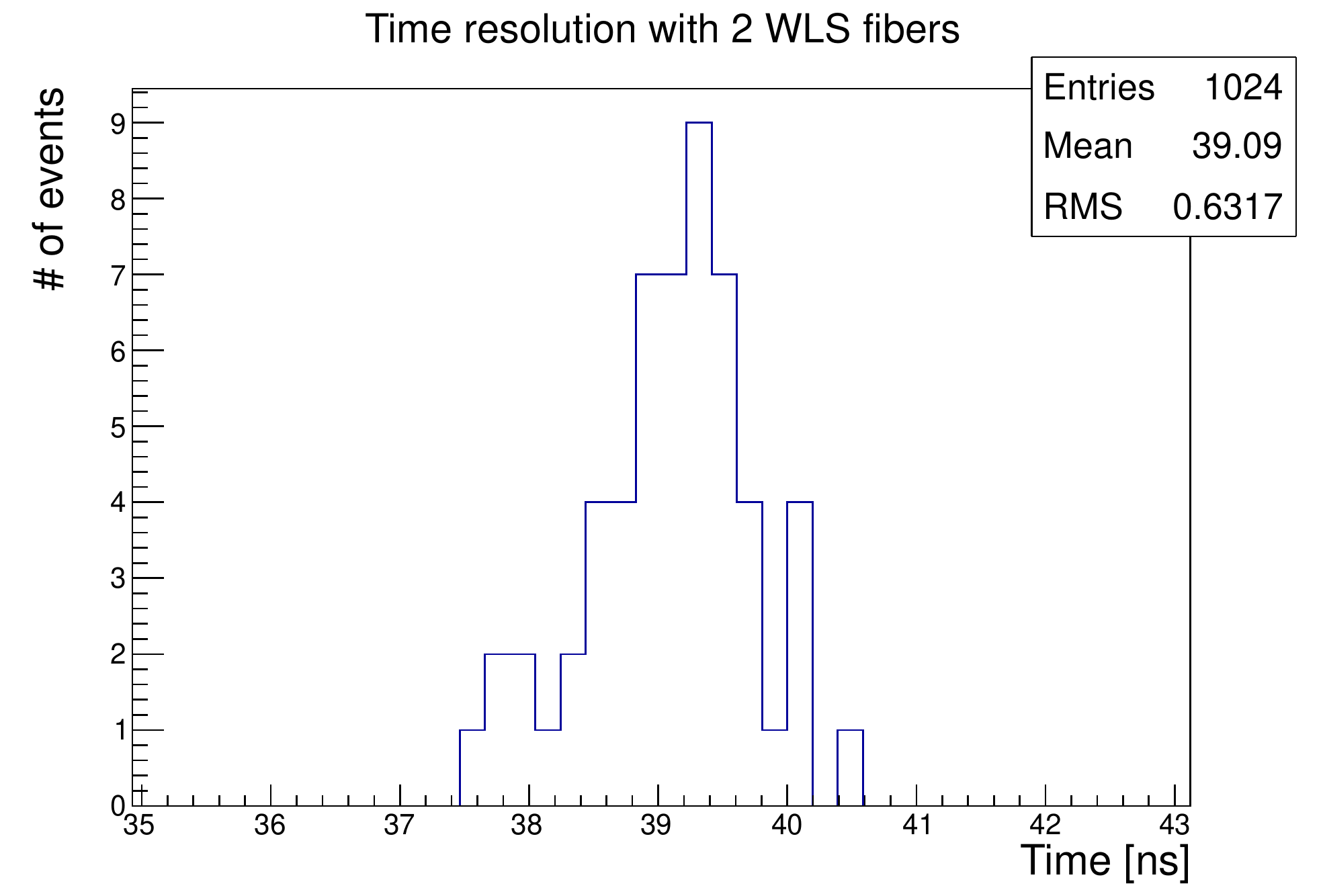} 
\caption{\label{fig:timeresolution}
The measured time distribution of the cosmic particle hit in one scintillator cube is
shown when the light yield is collected in one WLS fiber (left) and two WLS fibers (right). 
}
\end{figure}

In Figure~\ref{fig:timeresolution} the detector time resolution measured with the setup described above is shown.
Thanks to the high light yield provided by this geometry combined with a
``fast'' readout electronics, we achieve a time resolution of a single MIP particle hit 
of about 
0.91 ns % from RMS
%0.67 ns % from Gaussian fit
in a single WLS fiber 
and about 
0.63 ns % from RMS
%0.53 ns % from Gaussian fit
for the case when the light is collected by two orthogonal WLS fibers.
Further improvement can be obtained by considering the light collected simultaneously by all the three orthogonal WLS fibers.
A double-end readout, though more expensive given the doubled number of readout channels, 
might also improve the time resolution by approximately 40\%. 

% TIME RESOLUTION PLOTS

\section{Simulations}
\label{sec:simulation}

The full detector, corresponding to the parameters in Table~\ref{Table:parameters_detector}, has been simulated
with the GEANT4 software \cite{geant4}. 
The detector response was parametrized and the track reconstruction was performed,
for the time being, without pattern recognition. 
However if there are more tracks in a single event they must fulfill separation criteria in order to be reconstructed. 
The axes $X$, $Y$ and $Z$ define respectively the width, the height and the length of the detector. 
A magnetic field of 0.2~T has been simulated along $X$, in the same configuration as ND280.
About 230k neutrino interactions in the detector 
%(SuperFGD) 
were simulated with the GENIE software \cite{genie}
%for a total of $10^{21}$ protons on target (p.o.t.)
with the neutrino beam centered along the $Z$ axis 
and the energy spectrum expected at ND280.

\begin{figure}
\begin{center}
\includegraphics[width=7.5cm]{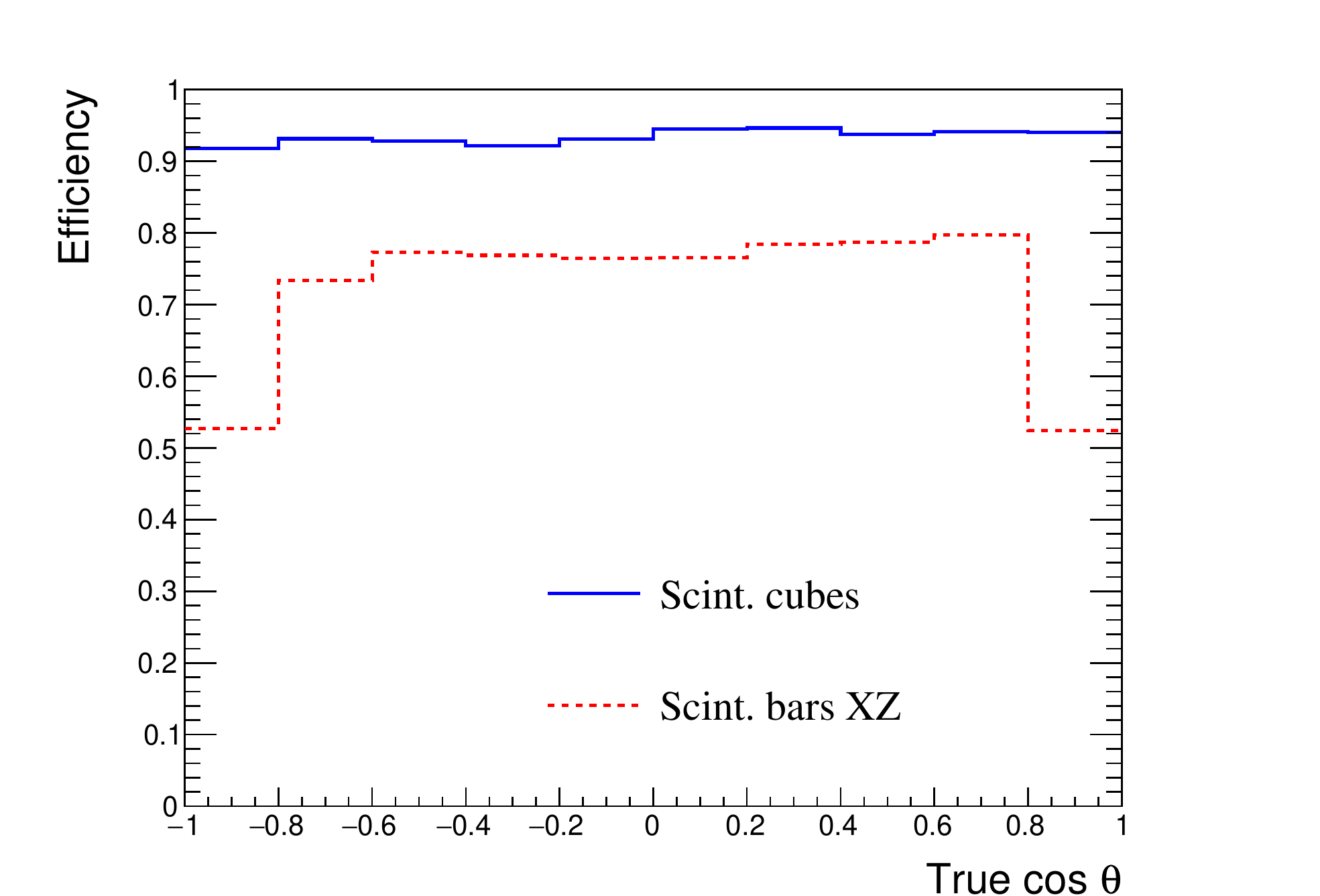} % TODO: update plot
\includegraphics[width=7.5cm]{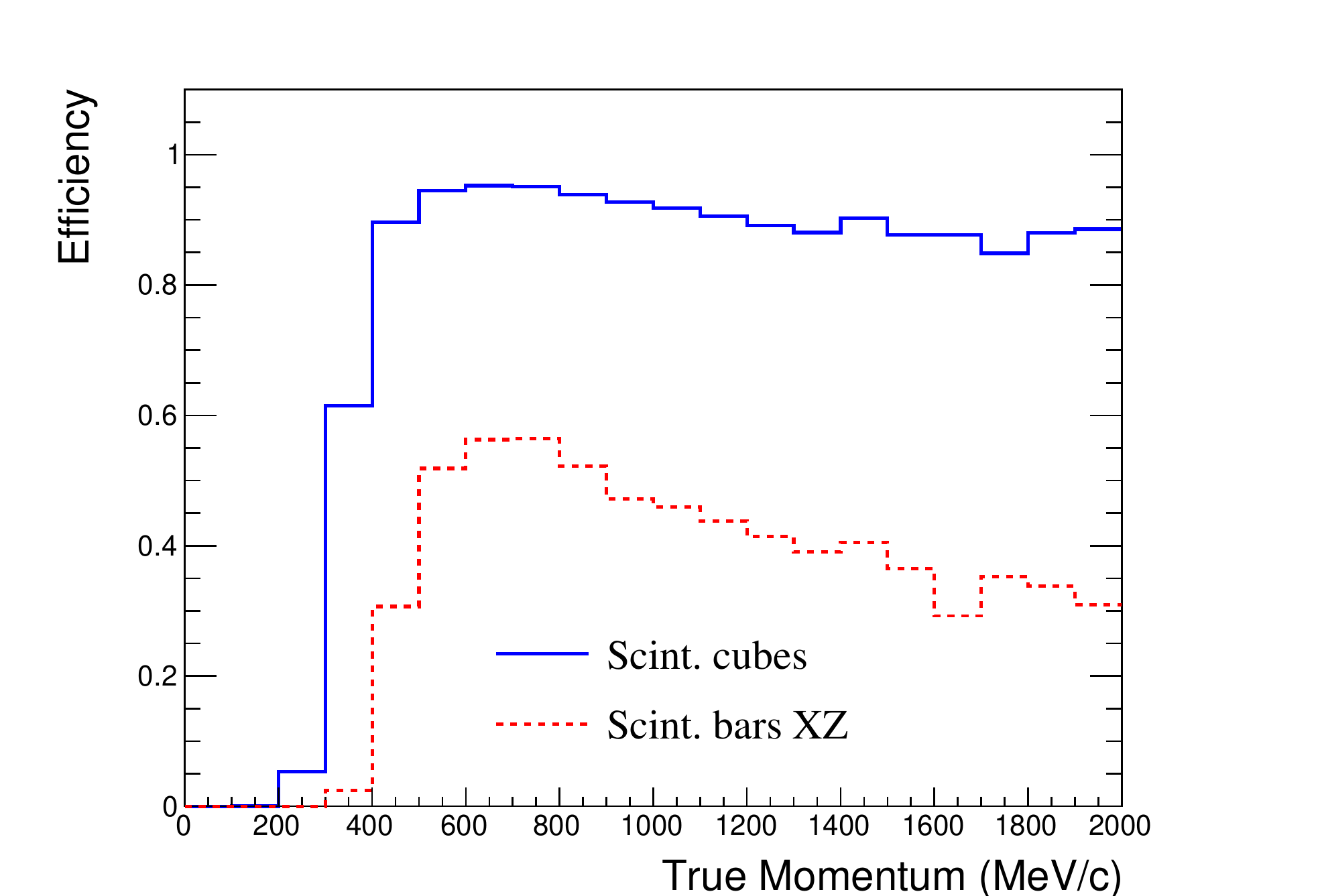}\\[6pt] % TODO: update plot
\caption{\label{eff_neutrino}
Track reconstruction efficiencies are shown for the particles produced by GENIE
neutrino interactions with $10^{21}$ p.o.t. 
for both the 3-views detector (named SuperFGD) proposed in this article and
a plastic scintillator detector made of bars along the X and Z directions (named FGD XZ). 
Left: muon reconstruction efficiency as a function of the truth muon $\cos \theta$.
Right: proton reconstruction efficiency as a function of the truth proton momentum.
 }
\end{center}
\end{figure}

The expected performance of the proposed detector was compared to a plastic scintillator detector made of 
$1 \times 1 \text{ cm}^2$ cross-section bars directed along the $X$ and $Z$ direction,
whose goal is to measure particles produced at about $90^{\circ}$ with respect to the neutrino beam direction.
% (FGD).
The results are shown in Figure~\ref{eff_neutrino}.
The reconstruction efficiency for muons is shown as a function of the muon angle with respect to the $Z$ axis.
It is clear that the detector proposed in this article has a $4\pi$ angular acceptance,
%with an efficiency always above 90\%. 
with a track reconstruction efficiency exceeding 90\% for the whole angular range.
Also the reconstruction efficiency as a function of the proton momentum is improved:
the momentum threshold to detect protons is reduced from about 450 MeV/c down to 300 MeV/c. 
This also shows that in the proposed detector it may be possible to improve the reconstruction of the neutrino
energy by measuring also low energy protons and pions.

We confirmed with a dedicated study that the particle identification capability is similar to that of the $XZ$ scintillator bars detector.  

\section{Conclusions}
\label{sec:conclusion}
We have shown that  the technique of extruded scintillator with wavelength shifting fiber readout can be extended to a three directional readout. 
The first simulations and 
the measurements with a small prototype of the conceived detector
%prototyping of the principle 
show encouraging results, with a light yield of more than 50 photo-electrons for one direction.  
A larger size prototype (aim is $5\times 5\times 5 \text{ cm}^3$ for a total of 125 cubes and 75 readout channels)  is under construction to be exposed to tests at particle beams, in which the timing properties and the possible cross-talk between channels can be evaluated. Further simulation studies will establish the predicted performance, for e.g. electron and photon separation or PID by  dE/dx, to be benchmarked at test beams. 
A challenging aspect of further design will be the mechanical integration providing sufficient rigidity as well as  compact and light disposition of the readout.  

 \section{Acknowledgements}
\label{sec:acknowledgements} 
This work was initiated in the framework of the T2K ND280 upgrade task force, convened by M. Yokoyama and M. Zito. Fruitful discussions in this context with our colleagues from T2K are gratefully acknowledged. 
D. Sgalaberna was supported by the grant number  200020-172709 of the Swiss National Foundation. 
The work  was supported in part   by the RFBR/JSPS  grant \# 17-52-50038.


\end{document}